# An amplitudes-perturbation data augmentation method in convolutional neural networks for EEG decoding

Xian-Rui Zhang
Department of Automation Sciences and
Electrical Engineering
Beihang University
Beijing, China
zhangxianrui89@163.com

Meng-Ying Lei
Department of Automation Sciences and
Electrical Engineering
Beihang University
Beijing, China
LMY_beginner@163.com

Yang Li*
Department of Automation Sciences and
Electrical Engineering
Beijing Advanced Innovation Center for
Big Data and Brain Computing
Beijing Advanced Innovation Center for
Big Date-based Precision Medicine
Beihang University
Beijing, China
liyang@buaa.edu.cn

*Abstract*—**Brain-Computer Interface (BCI) system provides a pathway between humans and the outside world by analyzing brain signals which contain potential neural information. Electroencephalography (EEG) is one of most commonly used brain signals and EEG recognition is an important part of BCI system. Recently, convolutional neural networks (ConvNet) in deep learning are becoming the new cutting edge tools to tackle the problem of EEG recognition. However, training an effective deep learning model requires a big number of data, which limits the application of EEG datasets with a small number of samples. In order to solve the issue of data insufficiency in deep learning for EEG decoding, we propose a novel data augmentation method that add perturbations to amplitudes of EEG signals after transform them to frequency domain. In experiments, we explore the performance of signal recognition with the state-of-the-art models before and after data augmentation on BCI Competition IV dataset 2a and our local dataset. The results show that our data augmentation technique can improve the accuracy of EEG recognition effectively.**

## I. INTRODUCTION

Electroencephalography (EEG) is an electro-physiological monitoring indicator that can be employed to analyze the states and activities of human brains. Traditionally, EEG is acquired by invasive ways, but recent years it is possible to collect EEG with noninvasive approaches [1]. So EEG-based Brain-Computer Interface (BCI) systems have gained popularity in various real-world applications from the healthcare domain to the entertainment industry. In healthcare domain, EEG signals are introduced to detect the organic brain injury or predict epileptic seizure [2]. Some BCI systems based on EEG have already allowing paralyzed patients to interact with wheelchairs or control a robot through their motor imagery EEG signals [3, 4]. Motor imagery EEG is a kind of signals collected when a subject imagines performing a certain action (e.g., closing eyes or moving feet) but does not make an actual movement [5]. As for the entertainment domain, the EEG signals have been applied to assisted living, smart home, and person identification etc.

Exploring an effective way to extract robust feature and classify EEG signals effectively is an significant work to applications of EEG. Traditionally, the process of EEG signals recognition consists of two stages [6]: a feature extraction stage, where meaningful information is extracted from the EEG recordings; a classification stage, where a decision is made from the selected features [7]. Traditional feature extraction methods mainly include frequency band analysis [8], multiscale radial basis functions [9], independent component analysis [10], continuous wavelet transform [11], and common spatial pattern algorithm [12] etc. In the classification stage, many traditional algorithms such as support vector machine (SVM) [13] and Bayesian classifier [14] have been employed. However, these methods heavily rely on handcrafted features, and the feature selection steps are time-consuming even for experts in this domain. Additionally, such a two-stage model is inconvenient to train and implement.

Recently deep learning techniques have gained widespread attention and achieved remarkable success in many fields such as computer vision [15], speech recognition [16], natural language processing [17] etc. Applying deep learning model to complete the feature extraction task show an appreciable performance as it does not require any handcrafting feature selection steps. Therefore, its effectiveness has encouraged some researchers to adopt deep learning methods to recognize EEG recordings. In [5], An et al. proposed a deep belief network (DBN) for classifying Motor imagery EEG and the DBN gained better result than the SVM method. In [14], Rezaei et al. adopted ConvNet to recognize EEG after converted EEG recordings to images by Short-Time Fourier

This work was supported by the National Natural Science Foundation of China [61403016, 61671042, and 61773039], Beijing Natural Science Foundation [4172037], and Open Fund Project of Fujian Provincial Key Laboratory in Minjiang University [MJUKF201702]. * Corresponding author.

*This manuscript has been accepted by ICCSS 2018.

Transform (STFT) [18]. More recently, Schirrmeister et al. proposed an end-to-end learning model called Shallow ConvNet which took raw EEG recordings as inputs [19]. It enable to learn robust feature representations and classify raw signals simultaneously. However, such an end-to-end ConvNet has more parameters to optimize than traditional methods, so we require a large amount of data to train this model.

For most public EEG datasets such as BCI Competition IV dataset 2a and 2b [20], there are only a few hundred samples per subject. Thus, if we try to further explore deep learning models on EEG recognition, applying a data augmentation technique is the primary issue that needs to be addressed [21]. Data augmentation is a technique that generate effective training samples from origin datasets when we do not have sufficient data to train deep learning model. In this paper we propose a augmentation method inspired from image processing technique. The augmentation approaches in images processing mainly include two parts, geometric transformations and noise addition. Geometric transformations, such as shift, scale, and rotation, are not practical for EEG as it is a dynamic time series. So we adopt the noise addition method to augment the EEG samples. However, before apply the noise addition method we must consider these characters of EEG recordings. First, EEG signals usually have low signal-to-noise ratio and a mass of noises such as eye blinks, muscle activity and heartbeat. Second, EEG signals lack sufficient spatial resolution compared to images [19]. Additionally, the key information of EEG signals mainly exists in the frequency domain [22]. Based on these knowledge of EEG, we modify the noise addition methods in image processing and implemented a novel data augmentation technique via adding Gaussian perturbation to the amplitudes in the frequency domain. Then we re-implement the Shallow ConvNet proposed in [19] and compare its performance with augmentation and without augmentation method on one public dataset and one local dataset. The experiment results demonstrate that our Gaussian perturbation method with proper parameters can yield significantly higher classification accuracy on both datasets.

## II. PROPOSED METHOD

In this section, we propose a data augmentation method called amplitudes-perturbation to improve the performance of the Shallow ConvNet model. In the following sub-sections we will describe the data augmentation method and Shallow ConvNet model in detail.

*A. Data Augmentation*

First we describe the notations in this paper. Each dataset is separated into labeled trials and each trial is a time segment of the original recording which belong to one of several classes. For trial $i$, the input sample of the model is represented as $X_i = \left[ X_1^c, X_2^c, \cdots, X_{t-1}^c, X_t^c \right]$ and the corresponding label is denoted as $y_i$. In sample $X_i$, each single 1-D vector $X_t^c$ contains $c$ elements acquired from all electrode channels at time index $t$. The input sample $X_i$ is a 2-dimension matrix, which can be arranged into an "image" with the number of discretized time steps $t$ as the width and the number of electrodes $c$ as the height.

Generally training a deep learning model needs a big amount of data otherwise the model tends to overfit quickly and is less robust. The data augmentation is a commonly used technique to generate samples from the existing dataset in deep learning because it is not time-consuming compared to collecting new data. The simplest data augmentation method is adding noise to the EEG signal in time domain directly. However, considering the characters of EEG such as low signal-to-noise ratio, non-stationarity and insufficient spatial resolution, this method may destroy EEG signal's feature in time domain [23]. So we propose the method that first transform the EEG recordings to the frequency domain via STFT [24], then add perturbations to amplitudes in frequency domain, finally reconstruct the time-series by inverse STFT.

Assume that we are given one training sample $X_i$ and its corresponded label $y_i$. First we denote the time-series of one channel as $x(t)$. Then transform it to frequency domain by short time Fourier transform as below:

$$Z(\tau, f) = \int_{-\infty}^{+\infty} x(t)g(t-\tau)e^{-j2\pi ft} dt \qquad (1)$$

where $Z(\tau, f)$ is a two-dimensional complex matrix representing the phase and amplitude of the signal over time $\tau$ and frequency $f$, and $g(t-\tau)$ is the Hann window function centered around zero. $Z(\tau, f)$ is a complex matrix so it can also be represented as $A\cos\varphi + jA\sin\varphi$, where $A \in \mathbb{R}^{f \times \tau}$ is the amplitude and $\varphi \in \mathbb{R}^{f \times \tau}$ is the phase in every frequency $f$ and time index $\tau$.

Then we add Gaussian-distributed noise $p \sim N(\mu, \sigma)$ to the amplitude randomly, where $\mu$ is the mean value and $\sigma$ is the standard deviation of Gaussian noise. Next, combine the disturbed amplitude and original phase to generate new complex matrix as:

$$Z'(\tau, f) = (A+p)\cos\varphi + j(A+p)\sin\varphi \qquad (2)$$

where $Z'(\tau, f)$ is the new complex matrix, and $A+p$ is the new amplitude after adding Gaussian noise.

After that, the new time-series $x'(t)$ is reconstructed by inverse STFT as:

$$x'(t) = STFT^{-1}(Z'(\tau, f)) \qquad (3)$$

where $STFT^{-1}$ represents the process of inverse Short-Time Fourier Transform.

Finally we apply all the above procedures to every channel in $X_i$, so we create a new perturbed training sample $X'_i$. Note that the new training sample has same target $y_i$ with the origin sample.

*B. Deep learning model*

In this paper, we use the Shallow ConvNet proposed in [19] to extract feature representations and classify the EEG signals automatically. The model consists of one temporal convolutional layer, one spatial convolutional layer, one meaning pooling layer, and one classification layer. In the first temporal convolutional layer, we utilize 25 convolution filters across time with a common kernel shape of (1,11) to capture temporal features. The choice of such a convolutional kernel will result in preserving the number of channels after the convolution operation and reducing the temporal dimension of the signals. But the temporal convolutional layer does not mix the channel signals with each other. In order to better handle the large number of input channels and mix the temporal representations of each channel, in the second spatial convolutional layer, 25 convolution filters of size (22,1) perform spatial filters with weights for all possible pairs of electrodes to learn spatial features. Usually the temporal features of each EEG channels are independent, so a common linear combination cannot be shared among the channels and a kernel size smaller than 22 is not ideal. The output of the spatial convolutional layer is the linear combination of all the 22 channels. Then the spatial feature maps are transmitted to the mean pooling layer with size (1,3) and stride (1,3). In the last classification layer, global convolution filters are applied to produce feature maps with size (1,1) and pass these feature maps to a softmax classifier directly, yielding the probability of the input belonging to each classes. So, in this layer, the number of filters and softmax units is equal to class labels.

Note that activation function between the temporal and spatial convolutional layers is not employed as it can regularize the overall neural networks in implicitly. Batch normalization, as recommended in [25], is applied to the output of the spatial convolutional layers before the exponential linear unit (ELU) function.

## III. EXPERIMENTS AND RESULT

In this section, we evaluate the performance of EEG signal recognition on the Shallow ConvNet model before and after data augmentation on BCI Competition IV dataset 2a and our local dataset. In the training stage of the model, it consists of two components: applying the amplitudes-perturbation data augmentation technique to the training dataset and feeding the augmented data to train the Shallow ConvNet. In the test stage, several typical metrics are employed to evaluate the performance with and without data augmentation technique, such as accuracy, precision, recall, F1 score, ROC (Receiver Operating Characteristic) curve, and AUC (Area under the curve). The Model Implementation, the result on public dataset, and the result on local dataset are separately reported in this section.

*A. Model Implementation*

We randomly select 80% samples from origin training set as new training set and the residual as validation set. All the dataset is divided into batches with the size of 64. The number of training iterations is set to 2000. The Adam Algorithms is adopted to minimize the cross-entropy loss function with a learning rate of 0.001. All convolutional layers have dropout with a probability of 0.5 [26].

*B. Result on BCI Competition IV dataset 2a*

The BCI Competition IV dataset 2a is a public EEG motor-imagery dataset including 9 subjects. The brain signals of each subject consist of two sessions which are recorded by 22 EEG electrodes according to the 10-20 electrode configuration. Each session consists of 288 trials of motor imagery tasks per subject. The movements of motor imagery include the left hand (class 1), the right hand (class 2), both feet (class 3), and the tongue (class 4). For each trial, EEG recordings are recorded at a sampling rate of 250Hz and low-pass filtered to 38 Hz. Subsequently we obtain trial epochs that starts at 0.5 s before the stimulus onset as input data and corresponding trial labels (class 1, class 2, class 3 and class 4) as targets. So we extract 288 training samples from the first session and 288 test samples from the second session each subject.

In this section, we report the performance study of Shallow ConvNet after amplitudes-perturbation data augmentation and then demonstrate the efficiency of our approach by comparing with the model trained without augmentation. As a first step before moving to the evaluation of data argumentation, we validate our Shallow ConvNet implementation. We reached an accuracy of 74%, statistically not significantly different from the origin research (73.7 %) in [19]. Then in the data augmentation technique, the mean value of Gaussian Noise is set to 0 in order to ensure the amplitude intensity not be changed. Then we set the standard deviation to 0.0001, 0.0005, 0.001, 0.002, 0.005 and 0.01 respectively, for purpose of exploring the effect of different standard deviations on the performance of Shallow ConvNet. Fig. 1 shows the recognition accuracies with Gauss noise of different standard deviations. The dashed line indicates the accuracy 74% without augmentation using Shallow ConvNet in [19]. From Fig. 1, we find that the standard deviation of Gaussian Noise can impact the performance of Shallow ConvNet effectively. If the standard deviation is too small or too large, we even get worse performance than the model without augmentation. When the standard deviation is set to 0.001, the model can achieve the best accuracy of 76.3% on four classes data.

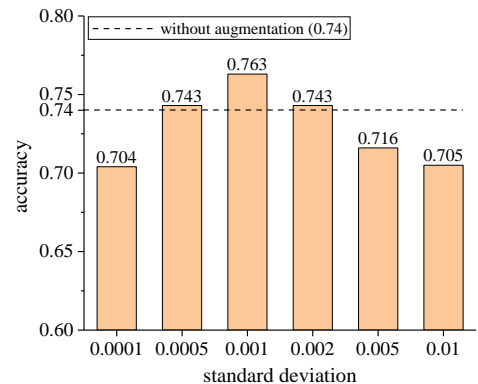

Fig. 1. The accuracy of Shallow ConvNet using data augmentation with different standard deviations. The dashed line indicates the accuracy of training without data augmentation.

In later experiments, we set the mean value to 0 and the standard deviation to 0.001 and note them as the best data augmentation parameters. In order to take a closer look at the result, the detailed confusion matrix, and other metrics of the model's performance are illustrated in Fig. 2, Fig 3 and Fig. 4 respectively.

In the confusion matrix of Fig. 2 and Fig. 3, the row

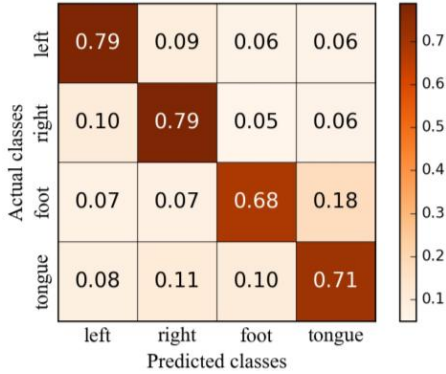

Fig. 2  Confusion matrix of training without data augmentation.

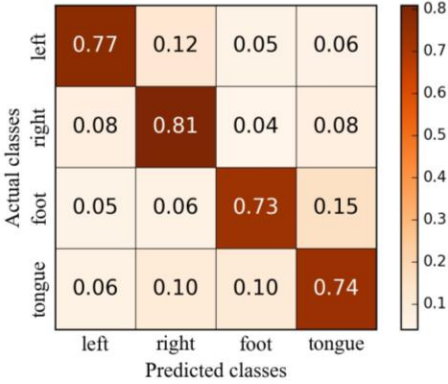

Fig. 3  Confusion matrix of training with data augmentation.

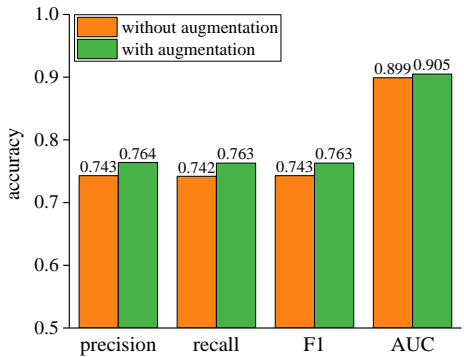

Fig. 4  Precision, recall, F1 and AUC of model with and without data augmentation.

represents the predicted classes $r$, the column represents the actual classes $c$. Each number in row $r$ and column $c$ denotes the ratio of target $c$ predicted as class $r$. The diagonal corresponds to correctly predicted trials of the four classes. Colors indicate fraction of trials in this cell from all trials of the corresponding row. The figure clearly depicts that for the classes of feet and tongue, our approach improve efficiently. In Fig. 4, we compare these two models in other metrics. It is observed that the performance can be improved obviously by our amplitudes-perturbation data augmentation method in precision, recall, F1 and AUC.

## C. Result on Local EEG Dataset

In this section, we evaluate the performance of Shallow ConvNet on our local dataset for demonstrating the good adaptability of the amplitudes-perturbation data augmentation technique. First, a brief description of the local dataset is given below. The local dataset is collected when the subjects are shown images of different scenery with the appearance time lasting 1s of one image. The shown images either contain an airplane (target), or no airplane(non-target). Participants are instructed to press a button when a target image is shown. The number of images with target and non-target is equal. When analyze the EEG signals, the P300 event-related potential is one of the strongest neural response to novel visual stimuli. So we can apply the Shallow ConvNet to recognize EEG trials with target from trials with non-target according to the P300 waveform. Compared to the public dataset BCI Competition IV 2a, our local dataset has two class labels and it is acquired from 64 channels at the sampling rate of 1000 Hz. We extract 320 training samples and 80 test samples from the local dataset. When evaluate on our dataset, the experiment setting and the parameters are same as we use on BCI Competition IV dataset 2a. Then we train the Shallow ConvNet with data augmentation technique and it reaches the accuracy of 88.75%, which is higher than the accuracy without using data augmentation (85.0%). The ROC curves of the two models are shown in Fig. 5. We can see that the curve with data augmentation is above the curve without data augmentation. So the AUC (area under the curve) value with data augmentation (0.93) is also higher than the value without data augmentation (0.91). It illustrates that the performance of the model with data augmentation is better than without data augmentation. In Table I, we list the precision, recall and f1 score of the model with and without augmentation. The improvement of the value with augmentation over without augmentation is obviously. It demonstrates that our data augmentation technique still achieves good performance and has good adaptability when it is applied to other dataset.

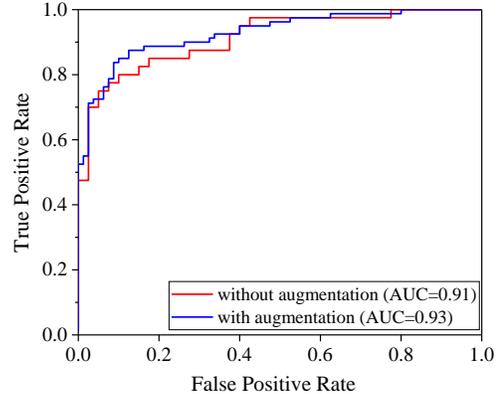

Fig. 5. ROC curves of the result with and without augmentation.

TABLE I. THE PRECISION, RECALL AND F1 SCORE OF THE MODEL WITH AND WITHOUT DATA AUGMENTATION

| method | precision | recall | F1 |
|---|---|---|---|
| without augmentation | 85.4% | 85.0% | 0.845 |
| with augmentation | 89.0% | 88.8% | 0.887 |

## IV. CONCLUSION

In this paper we present an amplitudes-perturbation method for EEG data augmentation. It yields significantly better performance on BCI Competition IV dataset 2a and our local dataset. First we explore the impact of standard deviation to the accuracy of Shallow ConvNet. Then we apply the best data augmentation parameters to train Shallow ConvNet and gain 2.3% higher accuracy than the model without amplitudes-perturbation on BCI Competition IV dataset 2a. Furthermore, the local dataset is employed to evaluate the adaptability of this proposal. The results show that the amplitudes-perturbation is a powerful method to improve the performance of deep learning models when training data is insufficient. Our future work will concentrate on improving the accuracy by other data augmentation methods, such as generative adversarial networks [27] and the variational autoencoder [28].